\documentclass[aps,prl,reprint,groupedaddress,nofootinbib]{revtex4-1}
\usepackage{hyperref}
\usepackage{amsmath}
 \usepackage{multirow}
\usepackage{array}
\newcolumntype{L}[1]{>{\raggedright\let\newline\\\arraybacksslash\hspace{0pt}}m{#1}}
\newcolumntype{C}[1]{>{\centering\let\newline\\\arraybackslash\hspace{0pt}}m{#1}}
\newcolumntype{R}[1]{>{\raggedleft\let\newline\\\arraybackslash\hspace{0pt}}m{#1}}
\usepackage{float}
\usepackage{graphicx}
\usepackage{epsfig}
\usepackage{psfrag}
\usepackage{color}
\usepackage{slashed}

\usepackage{amsfonts}
\usepackage{amssymb}
\usepackage{tikz}
\usepackage{tikz}
\usetikzlibrary{positioning,arrows}
\usetikzlibrary{decorations.pathmorphing}
\usetikzlibrary{decorations.markings}

\newcommand*{\be}{\begin{equation}}
\newcommand*{\ee}{\end{equation}}
\newcommand*{\bea}{\begin{eqnarray}}
\newcommand*{\eea}{\end{eqnarray}}

\newcommand{\comment}[1]{}

\newcommand{\cref}[1]{Chapter~\ref{c.#1}}

\def\beq{\begin{equation}}
\def\eeq{\end{equation}}
\def\bea{\begin{eqnarray}}
\def\eea{\end{eqnarray}}
\def\ba{\begin{array}}
\def\ea{\end{array}}
\def\bi{\begin{itemize}}
\def\ei{\end{itemize}}
\def\be{\begin{enumerate}}
\def\ee{\end{enumerate}}
\def\bc{\begin{center}}
\def\ec{\end{center}}
\def\bt{\begin{table}}
\def\et{\end{table}}
\def\btb{\begin{tabular}}
\def\etb{\end{tabular}}

\def\lsim{\raise0.3ex\hbox{$\;<$\kern-0.75em\raise-1.1ex\hbox{$\sim\;$}}}
\def\gsim{\raise0.3ex\hbox{$\;>$\kern-0.75em\raise-1.1ex\hbox{$\sim\;$}}}
	
\usepackage{cancel}
 \usepackage[normalem]{ulem}
\usepackage{color}
\def\comment#1{\textcolor{blue}{\large(\it{#1})}}

\def\lapp{\mathrel{\rlap{\raise.5ex\hbox{$<$}}
                    {\lower.5ex\hbox{$\sim$}}}}
\def\gapp{\mathrel{\rlap{\raise.5ex\hbox{$>$}}
                    {\lower.5ex\hbox{$\sim$}}}}
\begin{document}

\title{Testing the charge-radius coupling of composite Goldstone (Higgs) bosons at hadron colliders}

\author{G. Cacciapaglia$^{1,2}$}
\email{giacomo.cacciapaglia@in2p3.fr}
\author{S. Gascon-Shotkin$^{1,2}$}
\email{Susan.Shotkin.Gascon@cern.ch}
\author{A.Lesauvage$^{1,2}$}
\email{a.lesauvage@ipnl.in2p3.fr}
\author{N. Manglani$^{3}$}
\email{namrata201284@gmail.com}
\author{K. Sridhar$^{4}$}
\email{sridharkgita@gmail.com}

\affiliation{
$^1$University of Lyon, Universit\'e Claude Bernard Lyon 1,
F-69001 Lyon, France\\
$^2$Institut de Physique des 2 Infinis de Lyon (IP2I), UMR5822, CNRS/IN2P3,  F-69622 Villeurbanne Cedex, France\\
$^3$  Shah and Anchor Kutchhi Engineering College, Mumbai 400088, India\\
$^4$ School of Arts and Sciences, Azim Premji University, Sarjapura, Bengaluru 562125, India}

\begin{abstract}
We explore the collider relevance of a charge-radius coupling among light mesons in composite Higgs models. In particular, we focus of a coupling of the photon to the composite Higgs and a composite singlet, arising from isospin violation in the underlying theory. This coupling offers a deep probe of the composite nature of the Higgs mechanism, being sensitive to the electromagnetic and weak isospin structure of its constituents. The main collider effect consists in the production of the Higgs boson in association with a light composite pseudo-scalar. 
We present an exploratory cut-and-count analysis at hadron colliders, like the LHC, showing that an efficient background suppression can be achieved. More sophisticated techniques, however, are necessary to select a sufficient number of signal events, due to the small production rates. This justifies further investigation of this channel, which is highly complementary to other searches for compositeness in the Higgs sector.
\end{abstract}

%\pacs{73.21.Hb, 73.21.La, 73.50.Bk}
\maketitle

%\section{Introduction}
%\comment{Suppressing Section}
\noindent 
%\section{Introduction}

The idea that the sector responsible for the breaking of the electroweak symmetry may be composite~\cite{Weinberg:1975gm,Dimopoulos:1979es,Eichten:1979ah} is as old as the Standard Model (SM) itself~\cite{Weinberg:1967tq}. The first theories were inspired by quantum chromodynamics (QCD), and as such they did not feature a light degree of freedom associated to the Higgs boson~\cite{Higgs:1964pj}. However, it was soon realised that enlarging the symmetry breaking pattern can generate a light Higgs boson as long as the unbroken subgroup is large enough to contain the electroweak symmetry $SU(2)_L \times U(1)_Y$. In this new version, it is the misalignment of the vacuum of the theory with respect to the weak gauging that breaks the symmetry~\cite{Kaplan:1983fs}. The light Higgs boson, therefore, emerges as a pseudo-Nambu-Goldstone boson (pNGB). 

The most minimal model is based on the coset $SO(5)/SO(4)$~\cite{Agashe:2004rs} with its foundations in the idea of holography~\cite{Contino:2003ve}, i.e.~a conjectured duality between warped extra dimensional theories and four-dimensional strongly-interacting conformal field theories. Minimality here refers to the number of light composite scalars, which amount to the Higgs boson and the longitudinal polarisations of the $W^\pm$ and $Z$ bosons. However, for models based on an underlying gauge-fermion dynamics (like QCD), minimality, now defined in terms of the number of elementary constituents, always requires at least the presence of an additional light pNGB~\cite{Cacciapaglia:2014uja,Cacciapaglia:2020kgq}. The new minimal model, therefore, features the symmetry breaking pattern $SU(4)/Sp(4)$~\cite{Kaplan:1983sm,Galloway:2010bp} and it has 2 massive pNGBs: the composite Higgs boson $h$ and a pseudo-scalar $\eta$. Note that the same theory can be studied in the ``Technicolor'' limit, where the Higgs boson may emerge as a light (non-pNGB) resonance and the two light pNGBs act as Dark Matter candidates~\cite{Ryttov:2008xe}. In the composite Higgs limit, the singlet $\eta$ cannot be a Dark Matter candidate because it couples to the electroweak gauge bosons via the Wess-Zumino-Witten topological term~\cite{Wess:1971yu,Witten:1983tw}: the absence of a coupling to photons at leading order~\cite{Galloway:2010bp} makes the $\eta$ very difficult to detect at colliders~\cite{Cacciapaglia:2014uja,Arbey:2015exa,Craig:2018kne}. The most promising channels involve decays of heavy top partners~\cite{Bizot:2018tds,Benbrik:2019zdp,Cacciapaglia:2019zmj} or the Higgs boson itself~\cite{BuarqueFranzosi:2020baz}. The presence of the singlet pNGB has also been used to study strong order phase transitions at the electroweak scale~\cite{Bian:2019kmg,Xie:2020bkl,DeCurtis:2019rxl} and new mechanisms to generate an asymmetric Dark Matter in extended models~\cite{Cai:2019cow}.

In this letter, we want to focus on anomalous couplings of photons to the composite pNGBs, which may reveal the internal structure of the scalar fields. Couplings to photons have long been considered as typical signatures for compositeness in the electroweak sector. 
The new coupling reads~\cite{Hietanen:2013fya}:
\begin{equation}
\mathcal{L} \supset e \frac{d_{B}}{\Lambda^{2}} \sin \theta\left(\eta \overleftrightarrow{\partial}_{\mu} h\right) \partial_{\nu} F^{\mu \nu}\,,
\label{eq:coupling}
\end{equation}
and it is generated by the fact that the components of the neutral pNGBs $h$ and $\eta$ carry electromagnetic charges. Thus, following Ref.~\cite{Hietanen:2013fya}, we will refer to it as the {\it charge-radius coupling}. The coefficient $d_B$ can be associated to the masses of the spin-1 resonances once the underlying model is specified~\cite{Hietanen:2013fya}. The angle $\theta$ describes the misalignment of the vacuum, and can be expressed in terms of the electroweak scale $v_\mathrm{SM} = 246$~GeV and the pNGB decay constant $f$ as
 \begin{equation}
	  	\sin \theta=\frac{v_{\mathrm{SM}}}{f} \,.
	  	\label{eq:sintheta}
\end{equation}
This relation is universal, in the sense that it does not depend on the specific coset of the underlying model~\cite{Belyaev:2016ftv,Liu:2018vel}.

To better understand the physics behind the charge-radius coupling, we will focus on a specific realisation, based on a confining hyper-colour gauge group $Sp(2 N_c)$ with two Dirac hyper-fermions $U$ and $D$, transforming as the pseudo-real fundamental representation. As in the traditional Technicolor model, the left-handed components form a doublet of the SM $SU(2)_L$, while the right-handed ones form a doublet of the global $SU(2)_R$ (where the hypercharge is identified with the diagonal generator). In this way, custodial symmetry is automatically respected by the model~\cite{Georgi:1984af}. This set-up was first discussed in Ref.~\cite{Kaplan:1983sm}, and later reconsidered in the Technicolor~\cite{Ryttov:2008xe} and in the composite Higgs~\cite{Galloway:2010bp} limits. As already mentioned, the two are related by the misalignment angle $\theta$, so that $\theta = \pi/2$ corresponds to the Technicolor limit and $\theta \approx 0$ to the composite Higgs one~\cite{Cacciapaglia:2014uja}.
To connect to the lattice results in Ref.~\cite{Hietanen:2013fya}, we recall that in the Technicolor limit a global $U(1)_X$ remains unbroken, under which $h$ and $\eta$ form a charged scalar $\phi = (h + i \eta)/\sqrt{2}$. The charge-radius coupling in Eq.~\eqref{eq:coupling} can be written as
\begin{equation}
\mathcal{L} \supset i e \frac{d_{B}}{\Lambda^{2}}\left(\phi^{*}{\partial}_{\mu} \phi\right) \partial_{\nu} F^{\mu \nu} + \mbox{h.c.} \,,
\label{eq:coupling2}
\end{equation}
where this coupling matches Eq.~\eqref{eq:coupling} in the limit $\sin \theta = 1$. The coefficient is related to the masses of the spin-1 resonances $\rho$ as follows~\cite{Hietanen:2013fya}:
\begin{equation}
	 	\frac{d_{B}}{\Lambda^{2}}=\frac{m_{\rho_{U}}^{2}-m_{\rho_{D}}^{2}}{2 m_{\rho_{U}}^{2} m_{\rho_{D}}^{2}}\,.
\end{equation}
The mass difference is generated by isospin violating effects in the theory, and is typically small, thus allowing us to approximate the coupling as
\begin{equation}
	  	\frac{d_{B}}{\Lambda^{2}} \approx \frac{\delta m_{\rho}}{m_{\rho}^{3}}\,.
\end{equation}
We remark that a similar coupling will also exist with the $Z$ boson replacing the photon, but with a different coupling strength.

In the following, we will focus on the phenomenology consequences of this coupling. Firstly, we observe that the amplitude generated by the coupling~\eqref{eq:coupling} vanishes for on-shell photons (or $Z$'s), thus it will not generate any decays of $h$ or $\eta$. Hence, the major effect will be in a novel production mechanism for the pNGBs, i.e. production of $\eta$ in association with a Higgs boson:
\begin{equation}
e^+ e^-,\; pp \to h\ \eta\,,
\end{equation}
via an off-shell photon or $Z$ boson. As in this letter we aim at testing the feasibility of this search, we will only include the photon coupling, knowing that the $Z$ can give additional contributions of the same order. A physically relevant case when the $Z$ contribution is negligible occurs in models where, in the underlying model, the operator in Eq.~\eqref{eq:coupling2} is generated for the hypercharge gauge boson. The mass of the pseudo-scalar $\eta$ can acquire very different values, depending on the details of the underlying models. In principle, the $\eta$ can be much lighter then the compositeness or the electroweak scales as its mass comes from explicit breaking of the $U(1)_X$ symmetry mentioned above. For masses $m_\eta < m_Z$, the dominant constraint on this new state come from $Z$ decays via $Z\to \eta \gamma$, which will be tested with high accuracy at the Tera-Z run of future $e^+ e^-$ colliders~\cite{Cacciapaglia:2021aqz}. Henceforth, we will consider $m_\eta > m_Z$ in the following: for $m_Z < m_\eta < 2 m_W$, the dominant decay is $\eta \to Z \gamma$; while for $m_\eta > 2 m_W$ the dominant decays will be $\eta \to W^+ W^-,\ ZZ$ \cite{Arbey:2015exa}. In this exploratory work we focus on the mass range $m_Z < m_\eta < 2 m_W$ because it has a unique final state with 100\% branching ratio, and because the presence of a monochromatic photon in the final state makes it much easier to extract from the background, as we will see later.

	\section{Production cross-sections}

	\begin{figure}[h!]
		\begin{center}
				\includegraphics[width=8 cm]{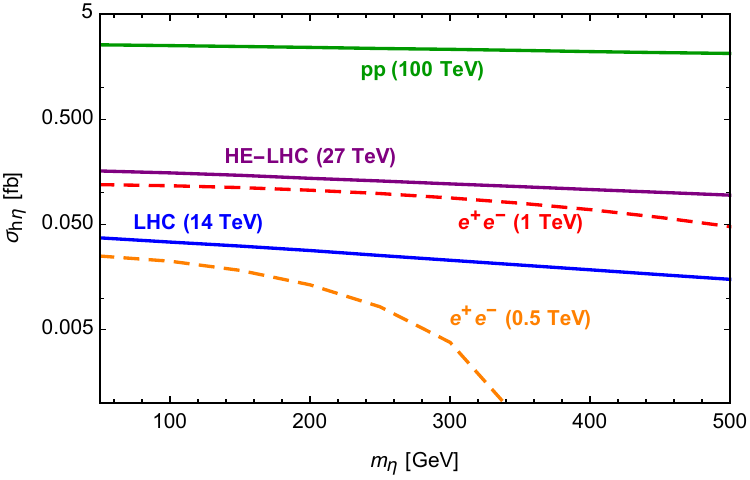}
		\end{center}
		\caption{\it Cross sections for the production of $h\eta$ via a virtual photon at various colliders. In solid colours, we show the LHC at 14 TeV, and other $pp$ future options. In dashed lines, future plans for $e^+ e^-$ colliders (ILC).}\protect\label{fig:xsec}
\end{figure}

We first present, in Fig.~\ref{fig:xsec}, the cross-sections for the process of concern at various colliders. First, the solid lines show the total cross-sections at hadron colliders, starting with the LHC run at 14 TeV in blue. The values correspond to the benchmark coupling
\begin{equation} \label{eq:bench}
\frac{d_B}{\Lambda^2} = 0.011~\mbox{TeV}^{-2}\,,
\end{equation}
corresponding to, e.g., $d_B = 0.1$ and $\Lambda = 3$~TeV. We recall that the cross-sections scale with the coupling squared. Note that the cross-sections are not very sensitive to the mass of the singlet $\eta$, thanks to the growth in energy of the amplitude, while increasing the centre-of-mass energy of the collider gives substantial gains: at 27 TeV, a factor of $\sim 6$ is gained, while it grows to roughly $\sim 100$ at 100 TeV.
In dashed lines, we also show the expected cross-sections at a future $e^+ e^-$ collider.

All in all, the cross-sections are extremely feeble, thus high integrated luminosities are necessary in order to constrain this signal. Moreover, as we will see, the backgrounds can be easily suppressed.
In the rest of this letter we will focus on estimating the potential reach of the High-Luminosity run of the LHC (HL-LHC), which is expected to collect $3$~ab$^{-1}$ of integrated luminosity, and use these results to project the potential reach of high centre-of-mass energy .

\begin{figure*}[tbh!]
	\begin{center} 
	\includegraphics[width=16cm]{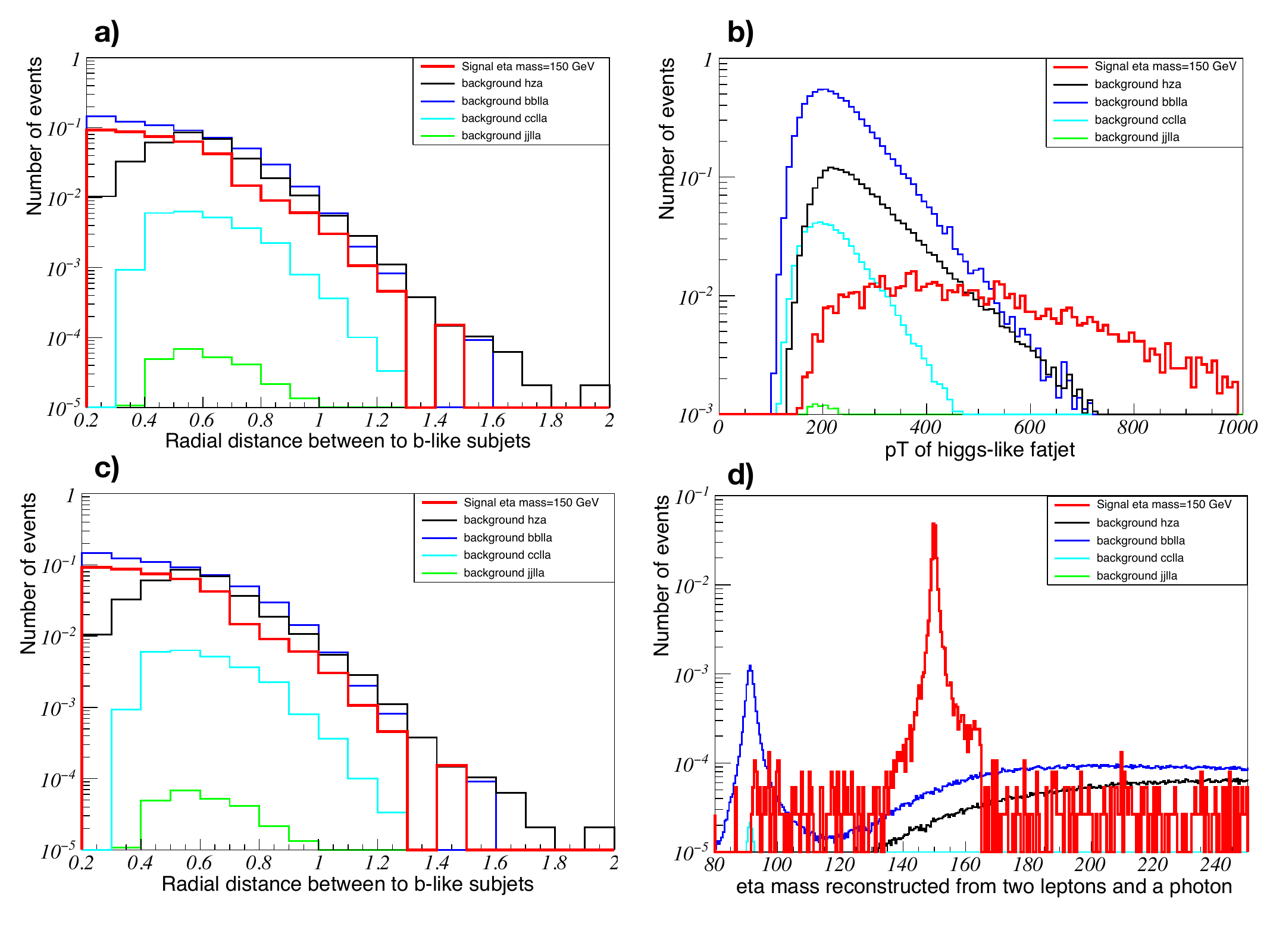}
	\end{center}
	\caption{\it Sample of kinematic distributions used in the cut flow. The top row shows characteristics of the Higgs-like fat-jet, the bottom row properties of the photon.} \protect\label{fig:kindist}
\end{figure*}

	\section{Signal and background simulation}
	
We want to focus on the following process:
\begin{equation}
pp \to \gamma^\ast \to h\eta \to h_{bb} Z_{ll} \gamma\,,
\end{equation}
at a centre-of-mass energy of 14 TeV, where we consider the Higgs boson decaying into $b$-jets and the $Z$ into leptons (electrons and muons). 
As an illustrative benchmark, we have chosen $m_{\eta}=150$~GeV in the simulations.
The model files for the $\eta$ couplings have been generated using {\tt{FEYNRULES}}~\cite{Alloul:2013bka}. The parton-level amplitudes for the signal and backgrounds were generated using {\tt{MADGRAPH}} \cite{Alwall:2014hca} for proton collisions at a centre-of-mass energy of $14$~TeV, using the parton distribution function {\tt{NNLO1}}~\cite{Ball:2012cx}. Showering and hadronisation were done in {\tt{PYTHIA 8}}~\cite{Sjostrand:2007gs}. 	 All cross sections and distributions we considered in this exploratory study are at leading-order. For the signal, we consider the benchmark coupling in \eqref{eq:bench}.
		
The SM provides an irreducible background $pp \to h Z \gamma$, which we will refer to as ``hza'' in the following. Additional backgrounds depend on the choice for the optimal decay channel for the Higgs and the $Z$ bosons. To maximise the signal yield, we choose the dominant Higgs boson decay, $h \to b\bar b$. For the $Z$, the cleanest channel is into two charged leptons (electrons and muons). While the branching ratio is small, this channel allows to reduce the pure QCD background and fully reconstruct the invariant mass of the $\eta$, which will be a crucial discriminant for the signal.  Hence, we will also include the irreducible background $pp \to l^+ l^- b \bar{b} \gamma$ (henceforth referred to as ``bblla''). To include the possibility of mistag of a light jet as a $b$-jet, we also simulate the reducible backgrounds $pp \to l^+ l^- c \bar{c} \gamma$ (henceforth referred to as ``cclla'') and $ l^+ l^- j j \gamma$ (henceforth referred to as ``jjlla''), where $j$ represents any light quark.
Another potentially relevant background is $t\bar{t} + \gamma$, with leptonic decays of the top quarks. However, it becomes 
negligible when we consider a boosted topology for the Higgs boson which decays to a pair of b quarks which are
more collimated, as compared to the $b$'s  from the tops which are almost back-to-back. 
In addition, we require two leptons from the top pair to reconstruct Z mass within a narrow window,
and the resultant Z to be closer to the photon. Hence, we will not include this background in the simulation.

\textbf{Event generation}: All events are generated with the minimal transverse momenta, $p_T$, of $40$~GeV for jets and $30$~GeV for leptons. Moreover, we define a signal region by selecting the following final state phase space: the invariant mass of the jet pair ranges from $30 < m_{b\bar b}< 200$~GeV; for the charged lepton pair $30 < m_{ll} <150$~GeV;  the maximum radial distance between two jets is $2$; the photon has a minimum transverse momentum of $p_T^\gamma >80$~GeV. 
The resulting cross-sections are reported in the first row of Table~\ref{tab:cutflow}, where the fiducial cross-sections in the signal region includes the branching ratios for the Higgs and $Z$ bosons. For the backgrounds, we generated $10^6$ events ($10^5$ for jjlla), while for the signal $10^4$ events.

\textbf{Higgs boson reconstruction}:
The final state particles, excluding photons and leptons, are clustered into fat-jets, i.e. a jet with large jet radius parameter, using  {\tt{FASTJET}}  \cite{Cacciari:2006sm,Cacciari:2011ma} by employing the Cambridge-Aachen algorithm \cite{Dokshitzer:1997in,Bentvelsen:1998ug}  for clustering, where we set the jet radius parameter $R = 1.2$.  We ensure that each event has two opposite sign leptons and a photon along with, at least, one fat-jet. The events which have a fat-jet with $p_T>50$ GeV are retained at the pre-selection stage.\\

We first define the criteria that identify the fat-jet as a Higgs boson candidate. We firstly ensure that it has two sub-jets, the radial distance between the sub-jets is less than $0.5$ (see Fig.~\ref{fig:kindist}a) and both sub-jets have a transverse momentum of $50$~GeV at least. The mass of the fat-jet is further required to be in a wide range around the Higgs mass, defined as $100 \leq M_{\rm fat-jet} \leq 170$~GeV. To better discriminate the signal from the background, we also require that the transverse momentum is greater than $380$~GeV, as illustrated in Fig.~\ref{fig:kindist}b. Note that the requirement on the angle between the two sub-jets is correlated to the high $p_T$ of the signal Higgs boson candidate with respect to the expected backgrounds. 

To control the dominant QCD background with light jets, we also impose that both sub-jets are tagged as b-jets. To simplify the analysis, we impose a flat $p_T$-independent probability on all the events. This probability is imposed at the end of the cut flow, described below, to retain the maximum number of events in the analysis. The working point we chose corresponds to the following probabilities: $70\%$ efficiency in tagging bottom quarks, with $20\%$ probability of mistagging a c-quark and $1\%$ for light jets \cite{CMS:2016kkf}.

\begin{table}[h!]
			\begin{center} 
			\begin{tabular}{|c|c|c|c|c|c|} \hline
 Cuts &  jjlla & cclla & bblla & hza & \phantom{x} $h\eta$ \phantom{x}\\ \hline
$\sigma_{\rm fid}$ (ab) & $590$ & $96$ & $93$ &  $19$ & $1.23$ \\ \hline
Events & $10^5$ & $10^6$ & $10^6$ & $10^6$ & $10^4$ \\
Higgs reco.& $337$ & $856$ & $3411$ & $4913$ & $1621$ \\	
$p_T^\eta > 380$ & $242$ & $695$ & $2572$ & $4625$ & $1566$ \\
$p_T^\gamma > 85$ & $231$ & $665$ & $2497$ & $4416$ & $1274$ \\
$\Delta R_{\gamma Z} <1.2$ & $57$ & $187$ & $636$ & $1260$ & $1259$ \\
$m_{ll}$ & $41$ & $128$ & $431$ & $1037$ & $1129$ \\
$m_{ll\gamma}$ & $2$ & $1$ & $9$ &  $29$ & $1085$ \\
\hline
b-tags & $0.0002$ & $0.04$ & $4.41$ & $14.2$ & $532$ \\
\hline
HL-LHC & $3.5 \cdot 10^{-6}$ & $1.2 \cdot 10^{-5}$ & $0.0012$ & $0.0008$ & $0.20$ \\
\hline
			\end{tabular}
		\end{center}
		\caption{\it Cut-flow for signal and backgrounds for the benchmark $\eta$ mass. We report the number of events in our simulation surviving after each requirement or cut. In the bottom row, we show the expected number of events at the HL-LHC with an integrated luminosity of $3$~ab$^{-1}$. For the cuts, the units are in GeV.}\label{tab:cutflow}
\end{table}

\textbf{Cut-flow analysis}: 
For our signal, the leptons and the photon originate from the decay of the $\eta$. As the Higgs boson has a substantial transverse momentum, the $\eta$ is also expected to be boosted, thus we could expect the angle between the photon and the $Z$ to be small. 
%Furthermore, the photon is expected to have a large transverse momentum itself. 
As shown in  Fig.~\ref{fig:kindist}c, imposing a maximum cut on the radial distance between photon and $Z$ allows to reduce the backgrounds, where the photon is dominantly radiated by the initial state partons. Hence, we impose a cut $p_T^\gamma > 85$~GeV and $\Delta R_{\gamma Z} < 1.2$. The cut on the photon transverse momentum allows to remove most of the background where the photon comes from soft radiation, at the price of reducing significantly the signal. 
The $Z$ is reconstructed by selecting a mass window for the lepton invariant mass, which we chose to be $86 \leq m_{ll} \leq 96$~GeV. 
Finally, we reconstruct the $\eta$ mass by the invariant mass of the photon and $Z$, and impose a cut in a narrow mass window around the mass of the pseudo-scalar (see Fig.~\ref{fig:kindist}d): $145 \leq m_{Z\gamma} \leq 156$~GeV. The cut-flow is summarised in Table~\ref{tab:cutflow}, with the number of events surviving each subsequent cut.  As already mentioned, the $b$-tagging efficiencies are imposed a flat rates, which we apply at the end in order to keep the full set of generated events. In the bottom row, we show the expected number of events at the HL-LHC, for an integrated luminosity of $3$~ab$^{-1}$ and for the benchmark value of the coupling.

\begin{figure}[h!]
		\begin{center}
				\includegraphics[width=8 cm]{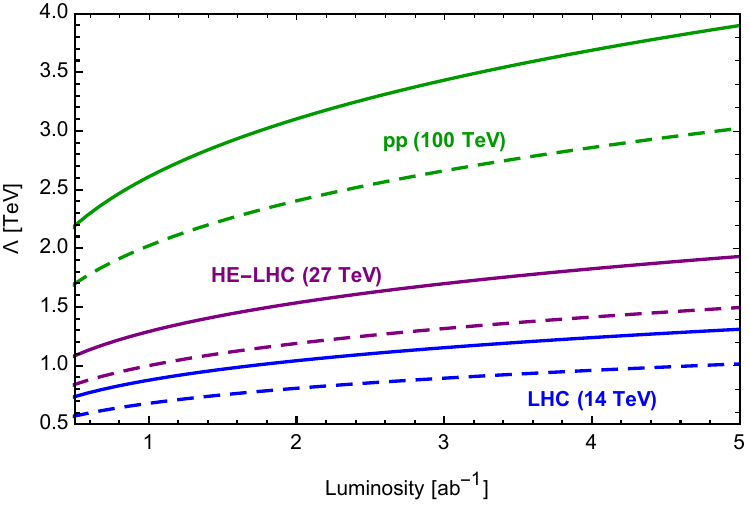}
		\end{center}
		\caption{\it Sensitivity to the scale $\Lambda$ (for $d_B = 0.1$) as a function of the integrated luminosity at various hadron collider energies. The solid lines indicate the $3\sigma$ reach, while the dashed ones correspond to $5\sigma$.}\protect\label{fig:sensitivity}
\end{figure}

\textbf{Results}:
The results of our proposed cut-flow demonstrate that the background can be significantly reduced compared to the signal. 
However, for the benchmark value of the coupling, no signal events survive at the HL-LHC due to the small production cross-section. Hence, we determine the sensitivity of the search by combining signal ($S$) and background ($B$) events as follows:
\begin{equation}
Z (\mathcal{L}, \Lambda) = \frac{S}{\sqrt{S+B}}\,,
\end{equation}
which depends on the coupling (in terms of $\Lambda$ for $d_B = 0.1$) and the integrated luminosity $\mathcal{L}$. We then use this formula to estimate the $3\sigma$ and $5\sigma$ reach in $\Lambda$ as a function of the luminosity, shown by the solid and dashed lines respectively in Fig.~\ref{fig:sensitivity}. For the HL-LHC, only scales up to the TeV can be probed. 

However, as we have seen in Fig.~\ref{fig:xsec}, the signal is highly enhanced for increasing energy of the proton collisions. 
We estimate the reach of the HE-LHC option at 27 TeV and of a future hadron collider at 100 TeV by rescaling signal and backgrounds by the ratio of the fiducial cross-sections, defined in the same fiducial region we used for the LHC. For the HE-LHC, the backgrounds are enhanced by a factor between $2.3$ and $3$, while the signal increases by a factor of $4.7$, thus allowing to probe scales $\Lambda$ up to $1.5$~TeV. For the 100 TeV collider, the enhancements amount to factors in the range $10\div 20$ for the background and $80$ for the signal. This allows to probe $\Lambda$ up to $3$~TeV, thus including our benchmark coupling. These projections are shown in purple and green, respectively, in Fig.~\ref{fig:sensitivity}.

\section{Concluding remarks}

We have presented a preliminary study of the collider relevance of the charge-radius coupling of a composite Goldstone Higgs boson, involving a photon and a singlet pseudo-scalar. The peculiarity of this coupling is that it entails the electromagnetic and isospin properties of the constituents of the composite Higgs boson, thus offering a direct probe of its nature. In particular, we focused on a mass range for the new pseudo-scalar where it decays dominantly to a photon and a $Z$ boson.

This coupling manifests itself in the pair production of the two spin-0 states, the Higgs and the pseudo-scalar bosons, via an off-shell photon. The rates, however, are very small, thus requiring high luminosities. We employed a simple cut-and-count strategy to reduce the reducible and irreducible backgrounds, proving that they can be suppressed enormously with respect to the signal, making this search a background-free one, effectively. However, the number of surviving signal events is very small. We show that the LHC in its high-luminosity phase can only access scales up to one TeV in the couplings, while a future 100 TeV collider can access up to $3$~TeV. This is due to the enhancement with energy of the signal, compared to the backgrounds.

This preliminary analysis demonstrates the feasibility of searching for this kind of signal from a deep probe of the composite nature of the Higgs boson. The sensitivity could be improved by designing a more effective procedure that allows more signal events to survive the selection. We believe that employing machine learning or deep learning techniques can help to achieve the desired improvement, as the signal has several kinematical features that help to distinguish it from the backgrounds. This will be the main objective of a future investigation.

\section*{Acknowledgements:}
We acknowledge the financial support of CEFIPRA on the project entitled ``Composite Models at the Interface of Theory and Phenomenology'' (Project No. 5904-C). 
GC is grateful to the LABEX Lyon Institute of Origins (ANR-10-LABX-0066) of the Universit{\'e} de Lyon for its financial support within the program ``Investissements d'Avenir'' (ANR-11-IDEX-0007) of the French government operated by the National Research Agency (ANR).

\bibliography{biblio}

%\newpage
%
%\widetext
%
%\appendix
%
%\section*{Supplementary material}
%
%\begin{figure*}[tbh!]
%	\begin{center} \begin{tabular}{cc}
%	\includegraphics[width=8cm]{Figures/ex150l_delr_bb.pdf} &
%	 \includegraphics[width=8cm]{Figures/ex150l_pth.pdf} \\
%	\includegraphics[width=8cm]{Figures/ex150l_delr_az.pdf} &
%	\includegraphics[width=8cm]{Figures/ex150l_ax_mass.pdf}
%	\end{tabular} \end{center}
%	\caption{\it Sample of kinematic distributions used in the cut flow. The top row shows characteristics of the Higgs-like fat-jet, the bottom row properties of the photon.} \protect\label{fig:kindist}
%\end{figure*}

\end{document}